\documentclass{elsart}
\usepackage[pdfborder={0 0 0}]{hyperref}  
\usepackage{graphics}
\usepackage{epsfig}
\usepackage{subfigure}
\usepackage{color}
\usepackage{amssymb}
\usepackage{amsmath}
\usepackage{float}

\journal{Physica A}

\begin{document}
\begin{frontmatter}

\title{Scaling laws in the diffusion limited aggregation 
of  persistent random walkers}
\author[irn]{Isadora R. Nogueira}, 
\author[ufv]{Sidiney G. Alves\corauthref{cor}}
\corauth[cor]{Corresponding author.}
\ead{sidiney@ufv.br}
\author[ufv]{Silvio C. Ferreira\thanksref{now}}

\thanks[now]{On leave at Departament de F\'{\i}sica i Enginyeria Nuclear, 
Universitat Polit\`ecnica de Catalunya, Barcelona, Spain.}

\address[irn]{Campus Alto Paraopeba, 
Universidade Federal de S\~ao Jo\~ao Del-Rei 
36420-000, Ouro Branco, MG, Brazil}

\address[ufv]{Departamento de F\'{\i}sica, 
Universidade Federal Vi\c{c}osa, 
36571-000, Vi\c{c}osa, MG, Brazil}

\begin{abstract}
We investigate the diffusion limited aggregation of particles executing
persistent random walks. The scaling properties of both random walks and large
aggregates are presented. The aggregates exhibit a crossover between ballistic
and diffusion limited aggregation models. A non-trivial scaling relation
$\xi\sim\ell^{1.25}$ between the characteristic size $\xi$, in which the cluster
undergoes a morphological transition, and the persistence length $\ell$, between
ballistic and diffusive regimes of the random walk,  is observed. 
\end{abstract}
\end{frontmatter}
\section{Introduction}

Aggregation processes are outstanding phenomena that have been subject of
theoretical, experimental and simulation investigations in many fields of
knowledge \cite{Meakin_Book,Sander}. The diffusion limited aggregation (DLA)
model proposed by Witten and Sander \cite{DLA} is the most studied theoretical
aggregation process. In this model, free particles are released, one at a time,
far from a growing cluster and perform successive random jumps while they
are not adjacent to the cluster. If a free  particle encounters any particle of
the cluster, it irreversibly sticks in the corresponding position and becomes
part of the cluster. Notwithstanding the rule simplicity, the DLA model produces
clusters exhibiting complex scaling properties
\cite{Mandelbrot,Ziff,SomfaiPhysa2003,Menshutin20086299}. Another simple and well
studied growth process is the so called ballistic aggregation (BA), in which the free
particles move ballistically at randomly chosen directions and obey the same
sticking rules as the DLA model.
{The BA model produces asymptotically homogeneous radial clusters
\cite{Alves_PRE} whose active (growing) zone is described by Kardar-Parisi-Zhang
(KPZ) universality class~\cite{KPZ}, as observed in the planar version where the
flux of particles is normal to an initially flat substrate~\cite{AaraoReis2006,Farnud}.}

Several generalizations of the DLA model have been studied \cite{Meakin_Book}
with special attention in those where the free particles perform random walks with
a drift \cite{Ferreira_PRE,Rodrigues_Perez,Castro2007,Huang}. In
these models, the clusters undergo a crossover from a DLA to a BA  scaling
regime as the number of particles increases due to the trajectories become
essentially ballistic at asymptotic large scales. In another group of models, in
which the free particles perform long random steps of a fixed length, an
inverted transition from BA to DLA is observed
\cite{Ferreira_EPJ,Alves_JCP}. In this case,  the trajectories are essentially
random at asymptotic large scales, independently of the step length. A simple
characterization of the morphological transition is done by relating the mass
$M$ of an aggregate with its radius $r$. {This quantity has two asymptotic
scaling regimes, $M\sim r^{2}$ and $M\sim r^{1.71}$, related to either the DLA
or BA models, respectively.}

{Aggregation models with drift or long steps can be used to investigate systems
with a long mean free path as, for example, the aggregation of 
methane or ammonia molecules in superfluid helium \cite{Vilesov2007}.
In particular, an off-lattice DLA model with long steps of length $\delta$ was
recently applied to the aggregation of molecules in superfluid media
\cite{Alves_JCP}. 
In addition to this example, biased diffusion
and super-diffusivity have also been reported in several systems ranging from
adatom surface diffusion \cite{Krugbook,Evans2006}  to cell migration
\cite{Dickinson,Kolega}. The relevance of modelling the aggregation with this
kind of random walk is strengthened due to biased diffusion may be misleadingly
confused with super-diffusivity \cite{ViswanathanPRE2005}.}

The representation of a large mean free path by a ballistic movement is a
simplifying hypothesis since weak perturbations can promote slight deviations of
the ballistic path resulting a biased diffusion at short scales. Motivated by
the wide availability of systems for potential applications, we investigate a
two-dimensional aggregation process where the particle trajectories consist of
random walks, with the direction of a new step limited by an angle
$\delta_\theta$ in relation to the previous step direction. This correlated
walk, which was formerly investigated in other contexts
\cite{Tojo,Bracher,Wu2000}, has a drift at short scales and becomes a random
walk at large scales.  Also, an aggregation model using this random walk was
formerly investigated in a small cluster size limit and a fractal dimension
depending on the $\delta_\theta$ parameter was reported~\cite{Huang}. Actually,
this dependence is a finite size effect since this kind of trajectory lead to a 
BA to DLA crossover, as demonstrated in the present work.

The paper is organized as follows. Model and simulation strategies are presented
in section \ref{model}. The trajectories are numerically investigated and an
analytical approach for the trajectory crossover is developed in section
\ref{sec:scaling}. 
In section \ref{results}, the
simulations of large clusters are presented and a scaling analysis of the
morphological transition is developed. Finally, some conclusions and prospects
are drawn in the section \ref{conclusions}.

\section{Model}
\label{model}

We perform  two-dimensional off-lattice simulations with an initial cluster
consisting of a single particle of diameter $a$ (a seed) stuck to the origin.
Free particles  are sequentially released, one at a time, at a circle of radius $r_l \gg
r_{max}$ centred at the seed, where $r_{max}$ is the largest distance from the
seed of a particle belonging to the cluster. The free particles follow a
persistent random walk \cite{Tojo}, in which the position of the $n$th step is
given by
\begin{eqnarray}
x_{n} =& x_{n-1} + a\cos\phi_n &\nonumber\\
y_{n} =& y_{n-1} + a\sin\phi_n & \label{trajectory}
\end{eqnarray}
and the direction of the $n$th step depends on the preceding one as 
\begin{equation}
\phi_{n}=\phi_{n-1} + \eta_n.
\label{eq:phin}
\end{equation}
The white noise $\eta_n$ is a random variable uniformly distributed in the interval
$(-{\delta_\theta}/2,{\delta_\theta}/2)$ and the parameter $\delta_\theta$
limits the next move direction inside an angular opening of size $\delta_\theta$ centred
in the direction of the previous step.

The trajectory becomes ballistic for $\delta_\theta\rightarrow 0$ and random for
$\delta_\theta\rightarrow 2\pi$ implying in the BA and DLA limit cases,
respectively. The direction of the first step  is chosen at random. The
trajectory is stopped whenever the particle visits a position adjacent to the
cluster where it irreversibly sticks. Finally, the particle is discarded
whenever it crosses a distance $r_k \gg r_{l}=r_{max}+\Delta$. The procedure is
repeated up to the cluster reaches $N$ particles. The previously introduced
variables $r_l$ and $r_k$ must be as large as possible, but computational
limitations restrict their values. For the DLA simulations, $\Delta$ can be a
few particle diameters~\cite{AlvesBJP}. However, for the BA limit, $\Delta$
cannot be small due to distortions caused by shadow instabilities
\cite{Tang,Yu}. The BA clusters are free from shadow effects when $\Delta>600$
and thus $\Delta=800$  and $r_k=10 \times r_l$ were adopted in all simulations.

\begin{figure}[t]
\begin{center}
\resizebox{7cm}{!}{\includegraphics*{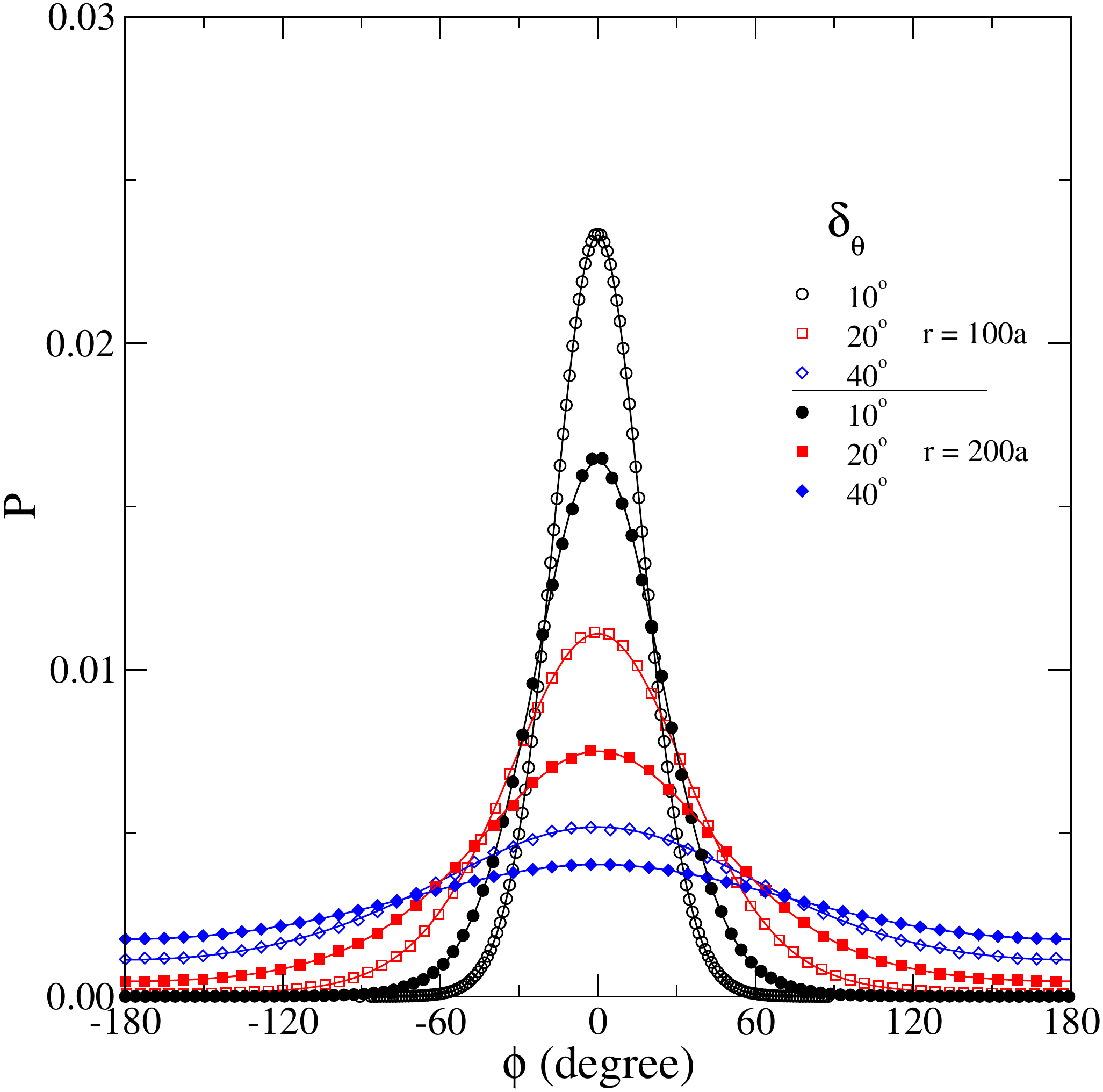}}
\end{center}
\caption{\label{ang_dist} (Colour on-line) Angular distributions of the first
passage at a distance $r$ for a persistent random walker starting at the origin.
Without loss of generality, we set $\phi_0=0$. Symbols represent the stochastic
simulations and solid lines non-linear regressions using Eq. \eqref{p_phifit}.}
\end{figure}

Since the asymptotic behaviour be typically reached only for large clusters, an
efficient algorithm is required. We adopted the  strategy where particles
execute long steps with the proper angular distribution $P(\phi)$ if they are
not close to the cluster \cite{Ferreira_PRE}. A closed expression for the angular
distribution of a long jump is not known a priori. So, we numerically computed
the probability of the first passage in a circle of radius $r$ occurring at an
angle $\phi$ for a trajectory started at the origin. Figure \ref{ang_dist} shows
some angular distributions for different values of $\delta_\theta$ and $r$
obtained with up to $10^8$  independent trajectories. If $r$ and $\delta_\theta$
are not too large, the distributions are very well fitted by 
\begin{equation}
P(\phi) = A \exp\left(B \cos\phi\right)+ C.\label{p_phifit}
\end{equation}
Parameters $A$, $B$ and $C$ (actually only two of them are independent due to
the normalization) were determined using non-linear regressions that are also shown in
Fig. \ref{ang_dist}. Cluster growth optimizations were implemented using steps
of size $16 a$ in the large empty regions nearby the cluster and three step
sizes ($16 a$, $100 a$ and $200 a$) were used far from the
cluster. More details about long step and off-lattice optimizations are
available elsewhere \cite{AlvesBJP}.

\section{Trajectory scaling properties}
\label{sec:scaling}

\begin{figure}[t]
\begin{center}
{\resizebox{4cm}{!}{\includegraphics*{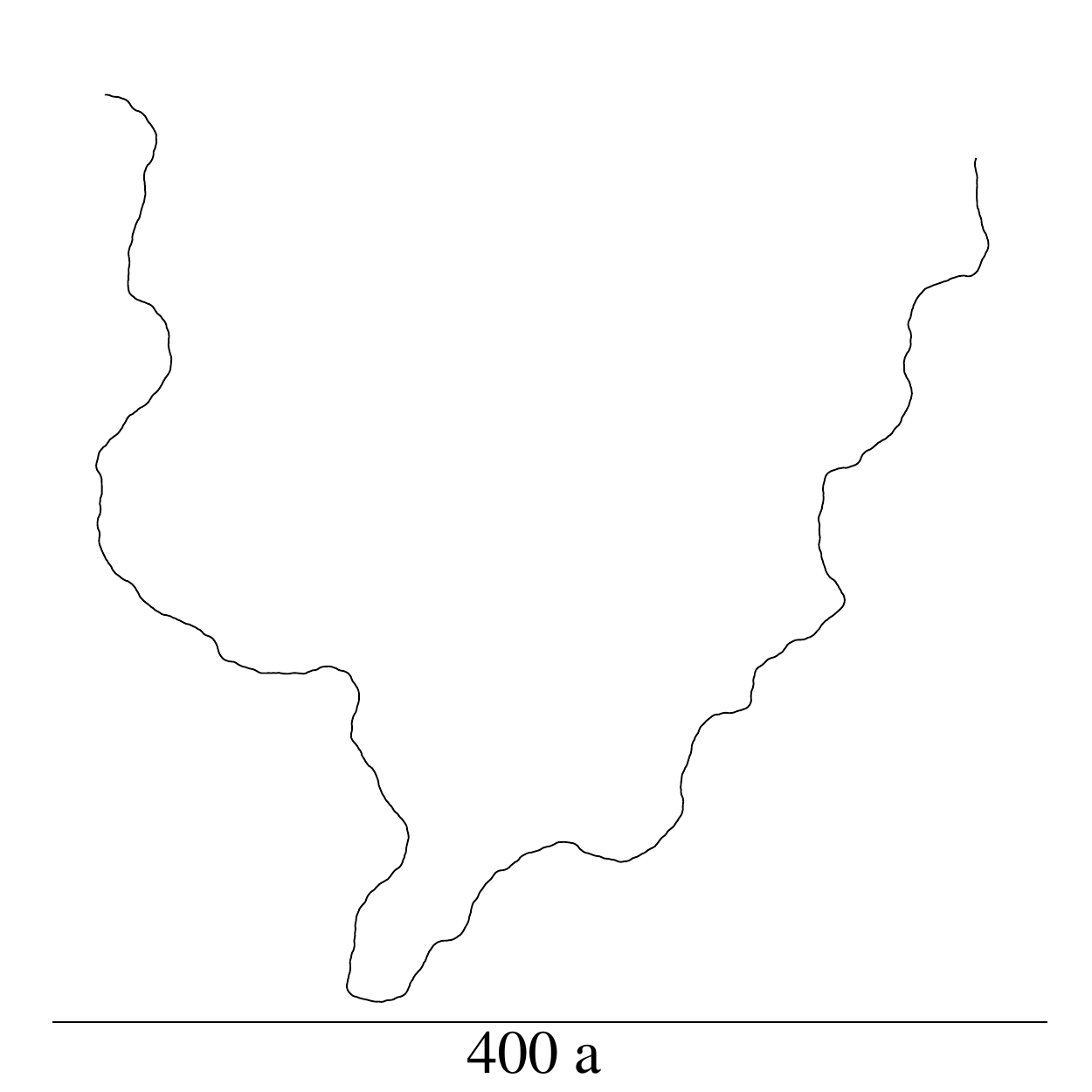}}}
{\resizebox{4cm}{!}{\includegraphics*{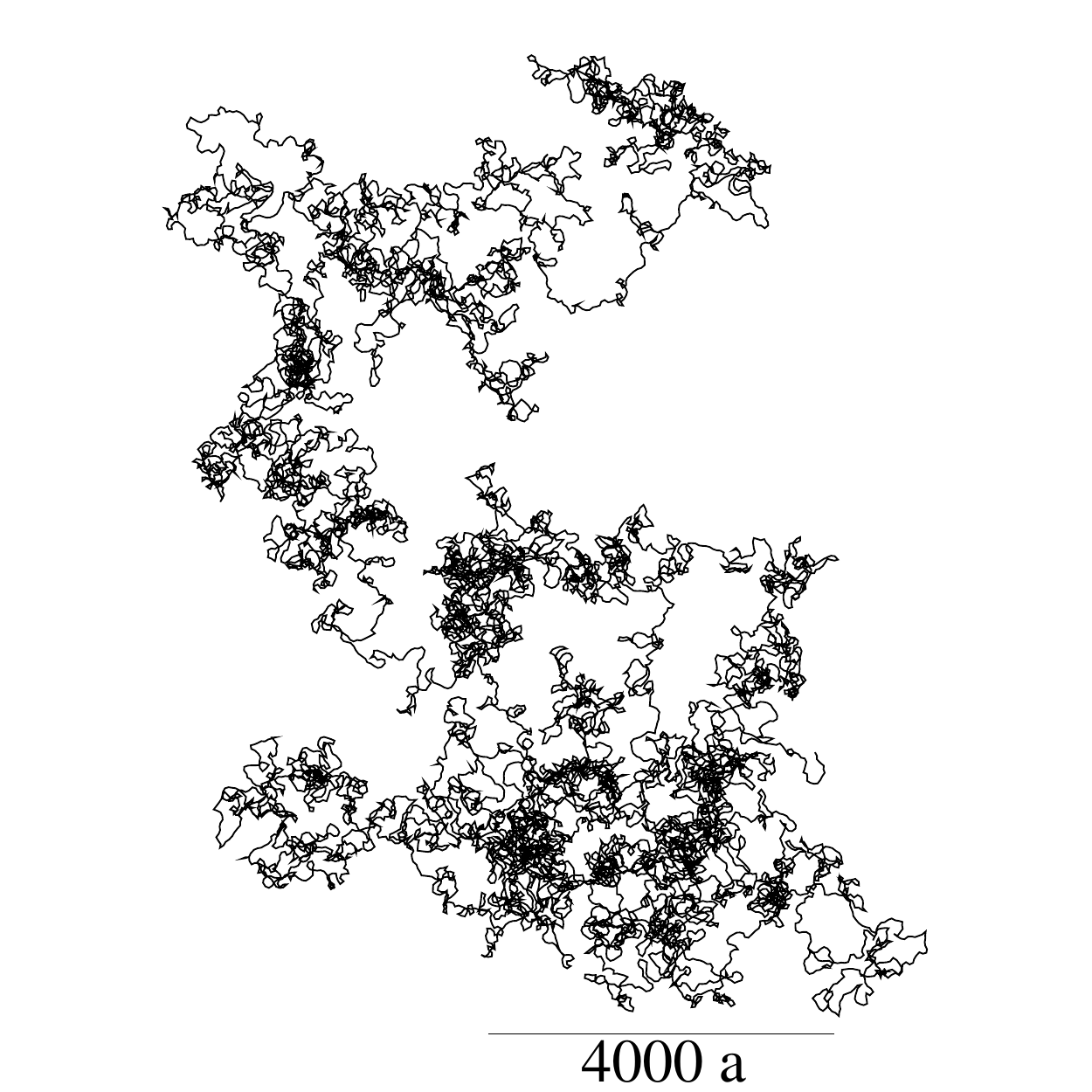}}}
\end{center}
\caption{\label{walks} A typical trajectory for 
$\delta_\theta = 10^\circ$ after $10^3$ (left) and $10^6$ (right) steps.}
\end{figure}

\begin{figure}[t]
\begin{center}
\subfigure[\label{rw_scaling}]{\resizebox{6.5cm}{!}{\includegraphics*{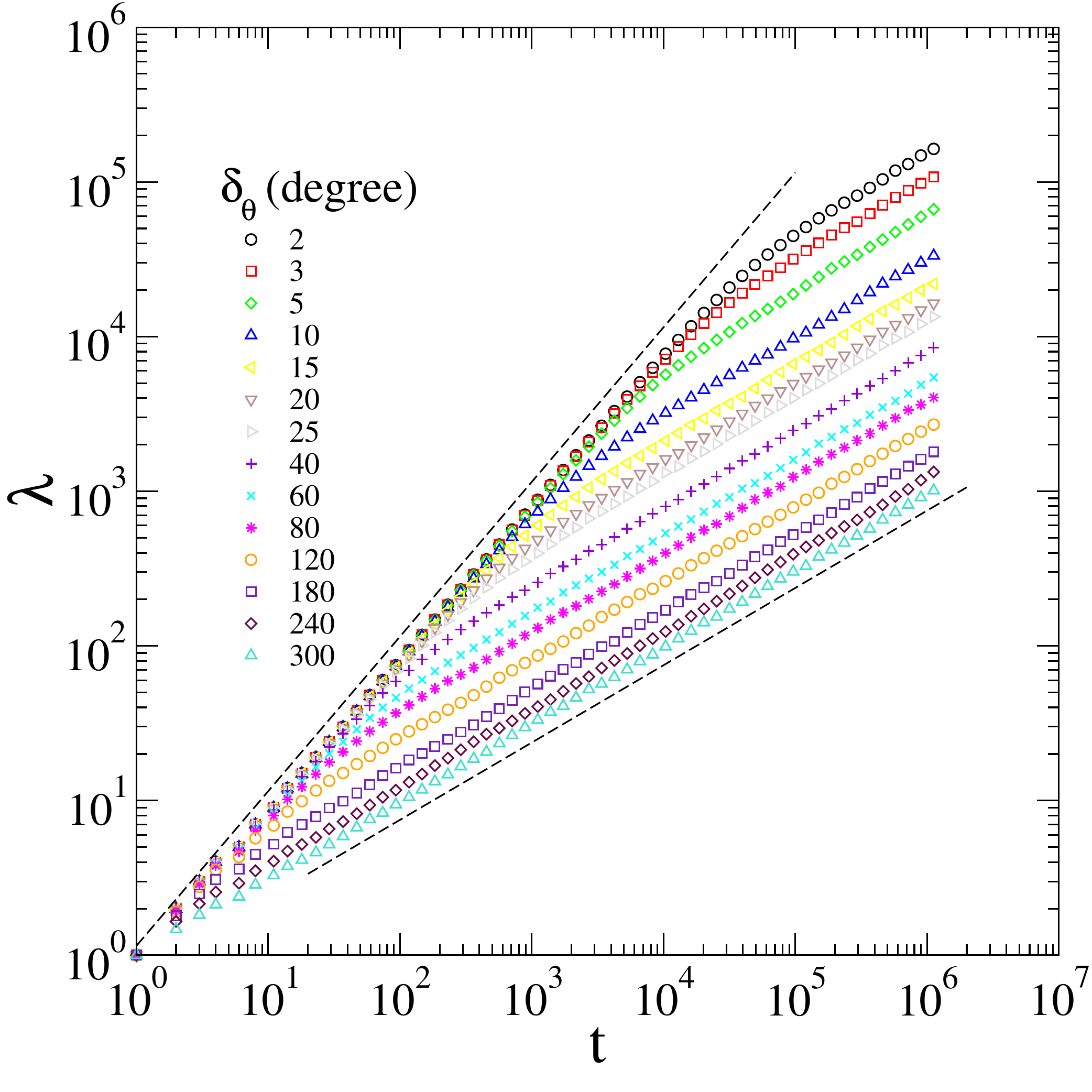}}}
\subfigure[\label{rw_collapse}]{\resizebox{6.5cm}{!}{\includegraphics*{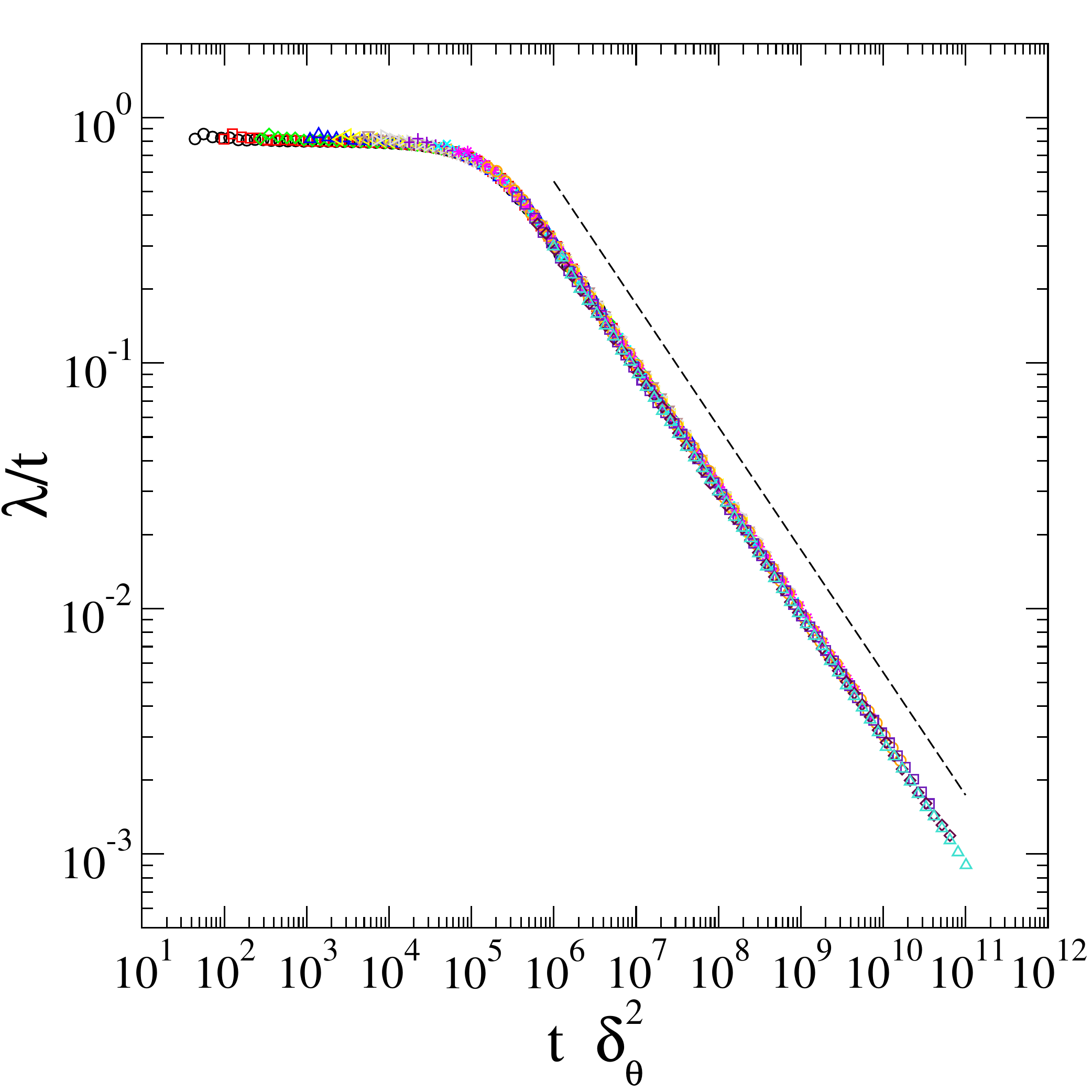}}}
\caption{(Colour on-line) (a) RMS displacements for distinct angular openings.
Dashed lines represent scaling laws $\lambda \sim t$ and $\lambda \sim t^{1/2}$.
(b) Collapse of curves shown in Fig. \ref{rw_scaling} using the scaling ansatz
given by Eq. \eqref{eq:lambda}. The dashed line represents the asymptotic
behaviour of the scaling function $g(x) \sim x^{-1/2}$.}                         
\end{center}
\end{figure}

Figure \ref{walks} shows two stages of a trajectory corresponding to short and long
times. Roughly, the trajectory consists in a drift around the initial direction
until a  large angular fluctuation be selected, as illustrated in the left panel
of Fig. \ref{walks}. The smaller angular opening the rarer the probability of a
large fluctuation be selected. However, these rare events unavoidably occur for a
sufficiently long time if $\delta_\theta\ne 0$ and the trajectories become
erratic at large scales, as shown in the right panel of Fig. \ref{walks}. In
order to investigate this crossover, $10^4$ independent trajectories were
simulated. In Fig. \ref{rw_scaling} the root-mean-square (RMS) displacement
$\lambda$ is show for several values of $\delta_\theta$. One can clearly resolve
a crossover: the RMS displacement scales as $\lambda \sim t$ at early stages
and crosses over to a scaling $\lambda\sim t^{1/2}$, as expected in a transition
from ballistic to diffusive trajectories.

The characteristic crossover time depends on the angular opening as $\tau\sim
\delta_\theta^{-2}$. This result can be proved with the following reasoning.
Without loss of generality, we assume $\phi_0=0$ in Eq. \eqref{eq:phin}
resulting \[\phi_t= \sum_{k=1}^{t}\eta_k,\] where $\eta$ is a white noise with 
$\langle\eta\rangle = 0$, $\langle\eta^2\rangle\propto\delta_\theta^2$, and 
$\langle\cdots\rangle$ means statistical averages. Obviously, the turning angle
is a random variable with $\langle \phi_t \rangle = 0$. Applying the central
limit theorem \cite{vankampen}, in which the sum of $t$ identical and
independent random variables with average $\langle \eta \rangle=0$  and variance
$\langle\eta^2\rangle$ converges to a normal distribution with variance
$\sigma_t^2 = t\langle\eta^2\rangle$, the probability of a large turning angle
be selected is exponentially negligible for $\sigma_t\ll 1$ and appreciable for
$\sigma_t\gg 1$. Assuming a crossover time given by \[\sigma_\tau =
\sqrt{\tau\langle\eta^2\rangle}\sim 1,\] we immediately find  out the scaling
law $\tau\sim\delta_\theta^{-2}$. Finally, the characteristic RMS displacement
at the crossover is proportional to $\tau$ since the trajectory is still 
approximately ballistic at the crossover. These scaling properties $\lambda$ 
can be expressed in the following scaling ansatz
\begin{equation}
\label{eq:lambda}
\lambda(t,\delta_\theta) = t \;\; g\left( \frac{t}{\tau} \right),
\end{equation}
where the scaling function $g(x)$ has the properties $g(x) \sim \mbox{const.}$ 
for $x\ll 1$ and $g(x) \sim x^ {-0.5}$ for $x\gg1$. As one can see in Fig.
\ref{rw_collapse}, this ansatz collapses the data shown in Fig. \ref{rw_scaling}
 onto a universal curve if we plot $\lambda / t$ against $t \times
\delta_\theta^2$, demonstrating the ansatz correctness. 

Before analysing the aggregates obtained using the persistent random walk, we
shortly draw a comparison with the results reported by Tojo and Argyrakis
for this persistent random walk
\cite{Tojo}, in which a scaling $\tau\sim \delta_\theta^{-1.88}$ was numerically
obtained in stochastic simulations. This exponent, that is slightly different
from ours, results from the crossover times estimated using the intersections
between early and long time scaling laws of $\lambda$ versus $t$. We instead
deduced the scaling behaviours and the theory perfectly matches the numerical
simulations as shown in Fig. \ref{rw_collapse}. 

\section{Aggregate simulations and scaling analysis}
\label{results}

\begin{figure}[t]
\begin{center}
\subfigure[~ $N = 10^5$]{\resizebox{2.75cm}{!}{\includegraphics*{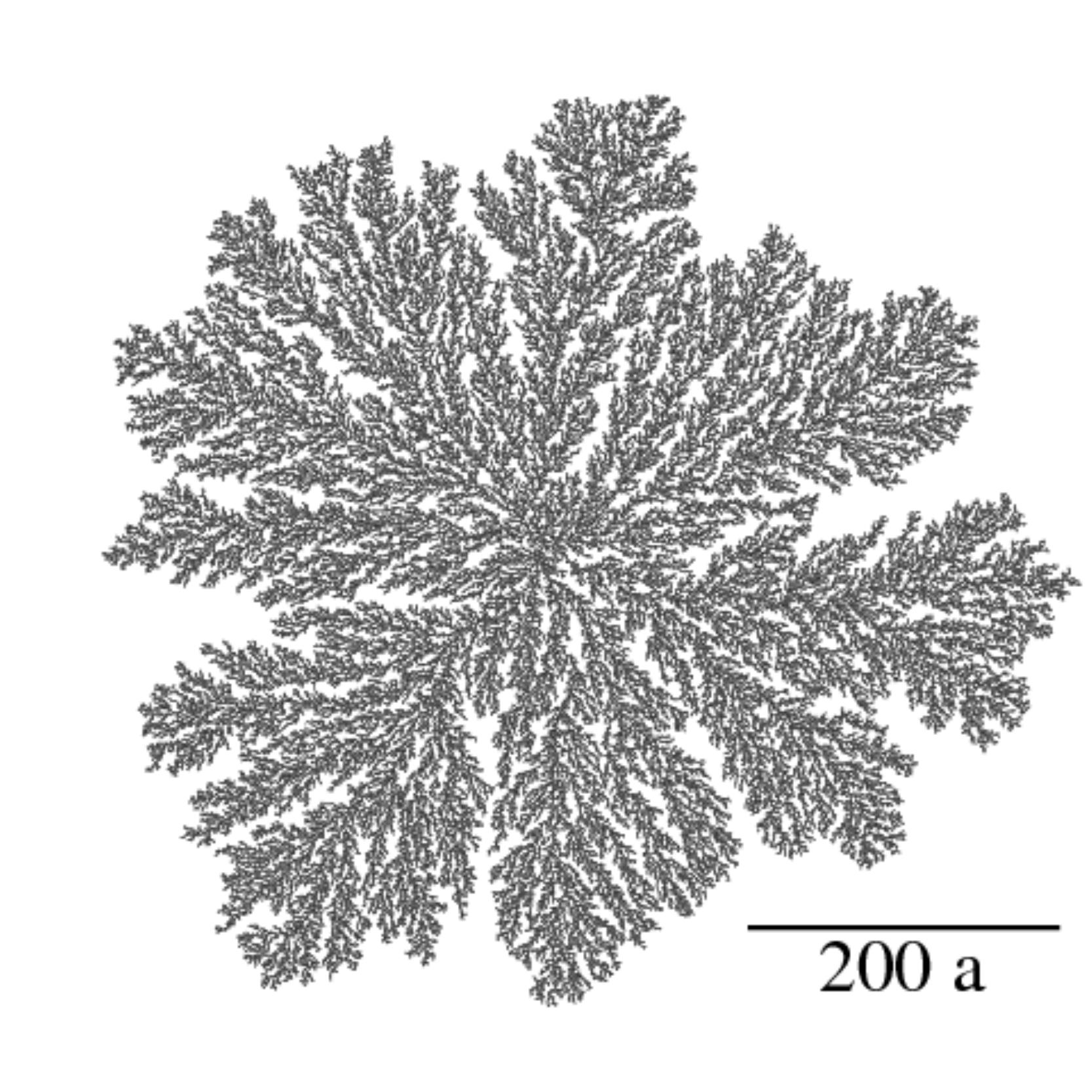}}}
\subfigure[~ $N = 10^6$]{\resizebox{2.75cm}{!}{\includegraphics*{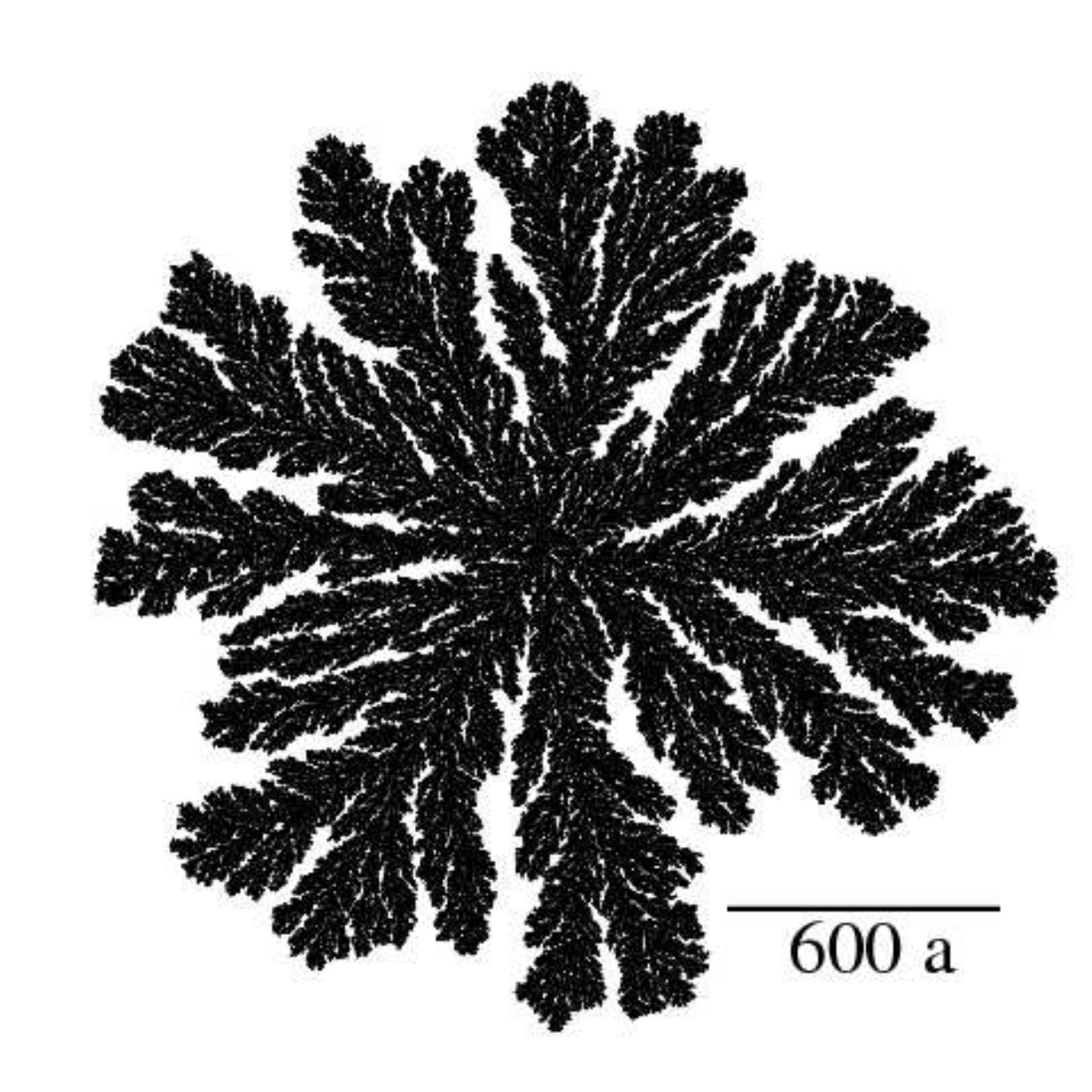}}}
\subfigure[~ $N = 3 \times 10^7$]{\resizebox{2.75cm}{!}{\includegraphics*{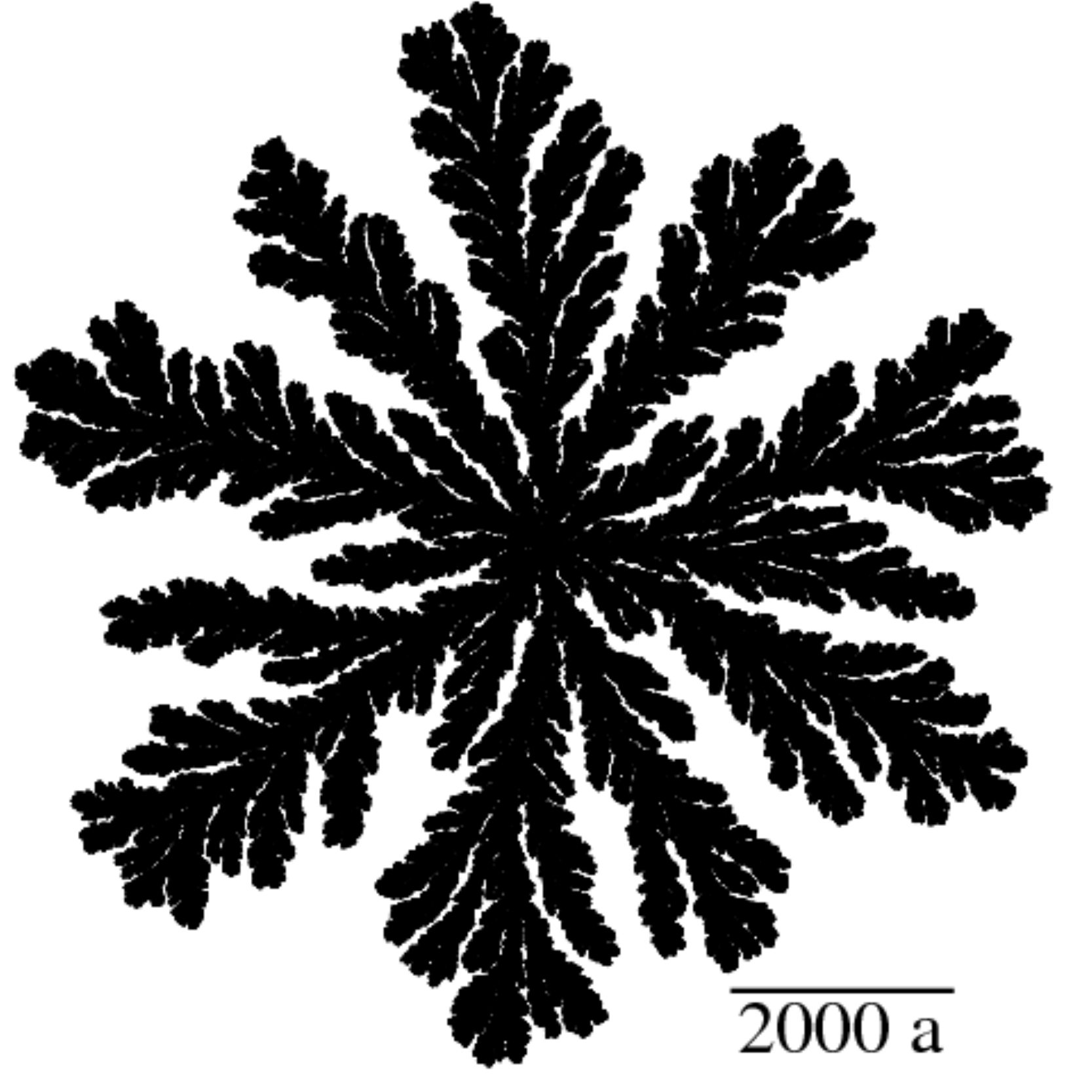}}}\\
\subfigure[~ $N = 10^5$]{\resizebox{2.75cm}{!}{\includegraphics*{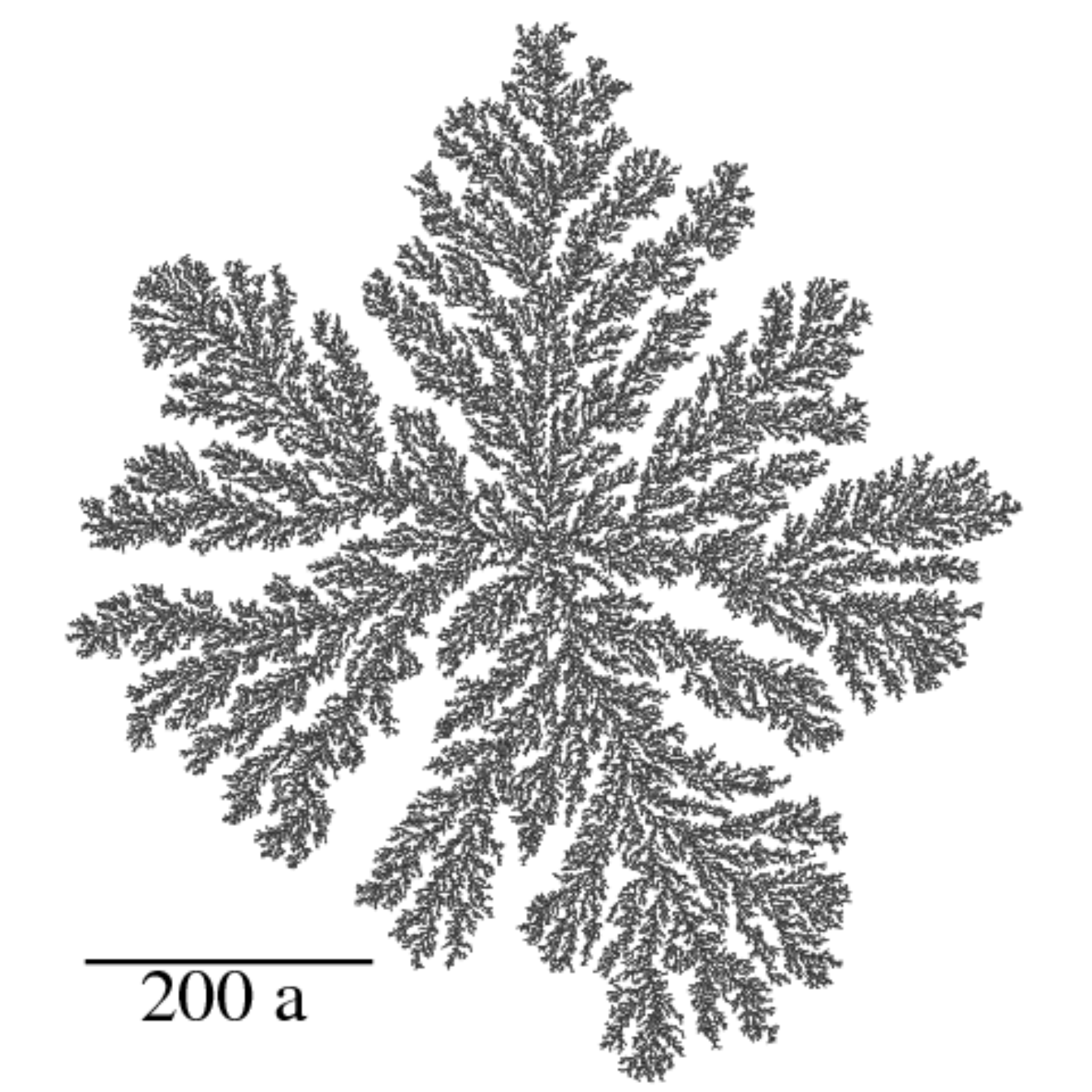}}}
\subfigure[~ $N = 10^6$]{\resizebox{2.75cm}{!}{\includegraphics*{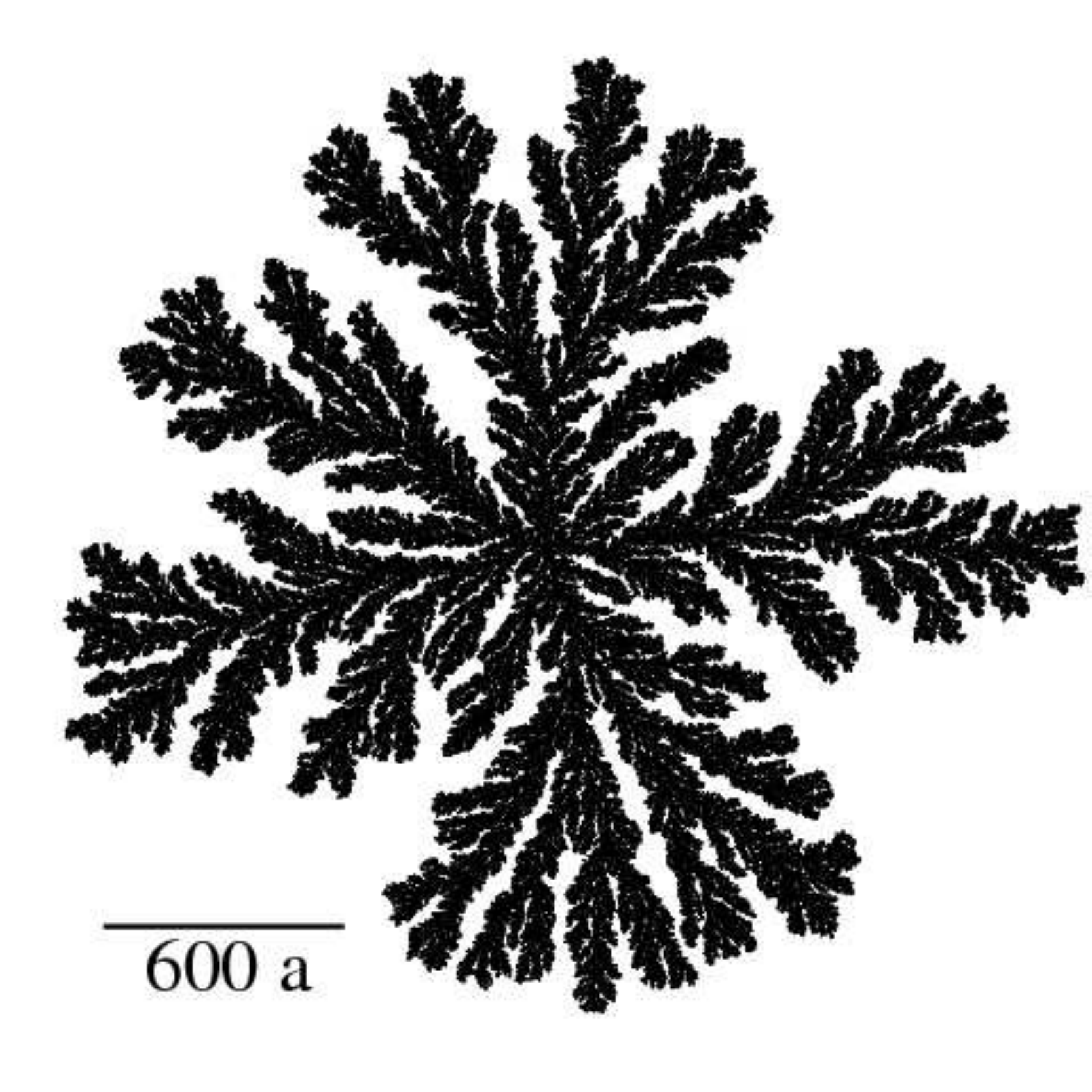}}}
\subfigure[~ $N = 10^7$]{\resizebox{2.75cm}{!}{\includegraphics*{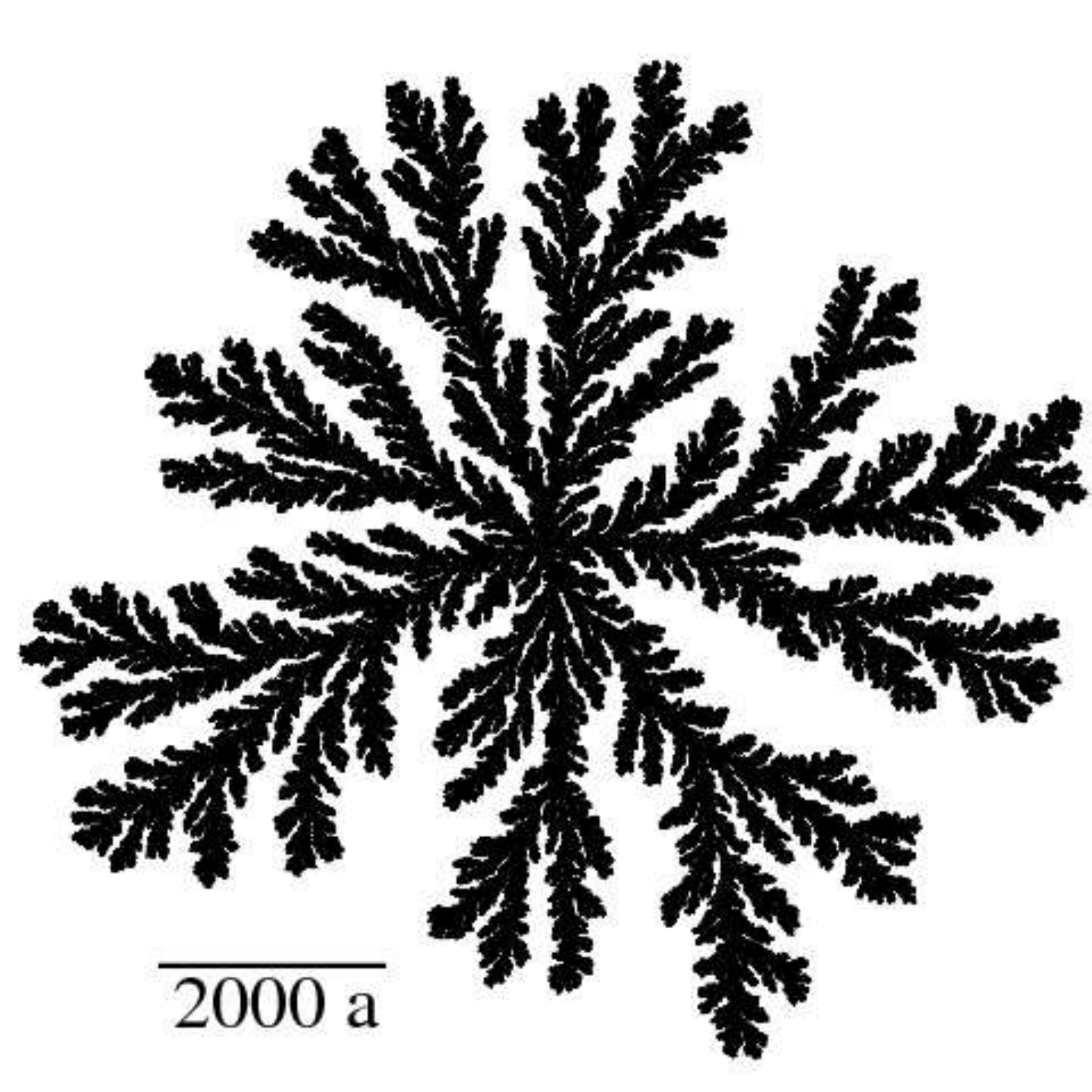}}}
\caption{\label{stages} Different growth stages of clusters generated for
$\delta_\theta = 40^\circ$ in (a)-(c) and $60^\circ$ in (d)-(f). The number of
particles are show below each panel.}
\end{center}
\end{figure}

\begin{figure}[t]
\begin{center}
\subfigure[~ $\delta_\theta = 10^\circ$]{\resizebox{3.2cm}{!}{\includegraphics{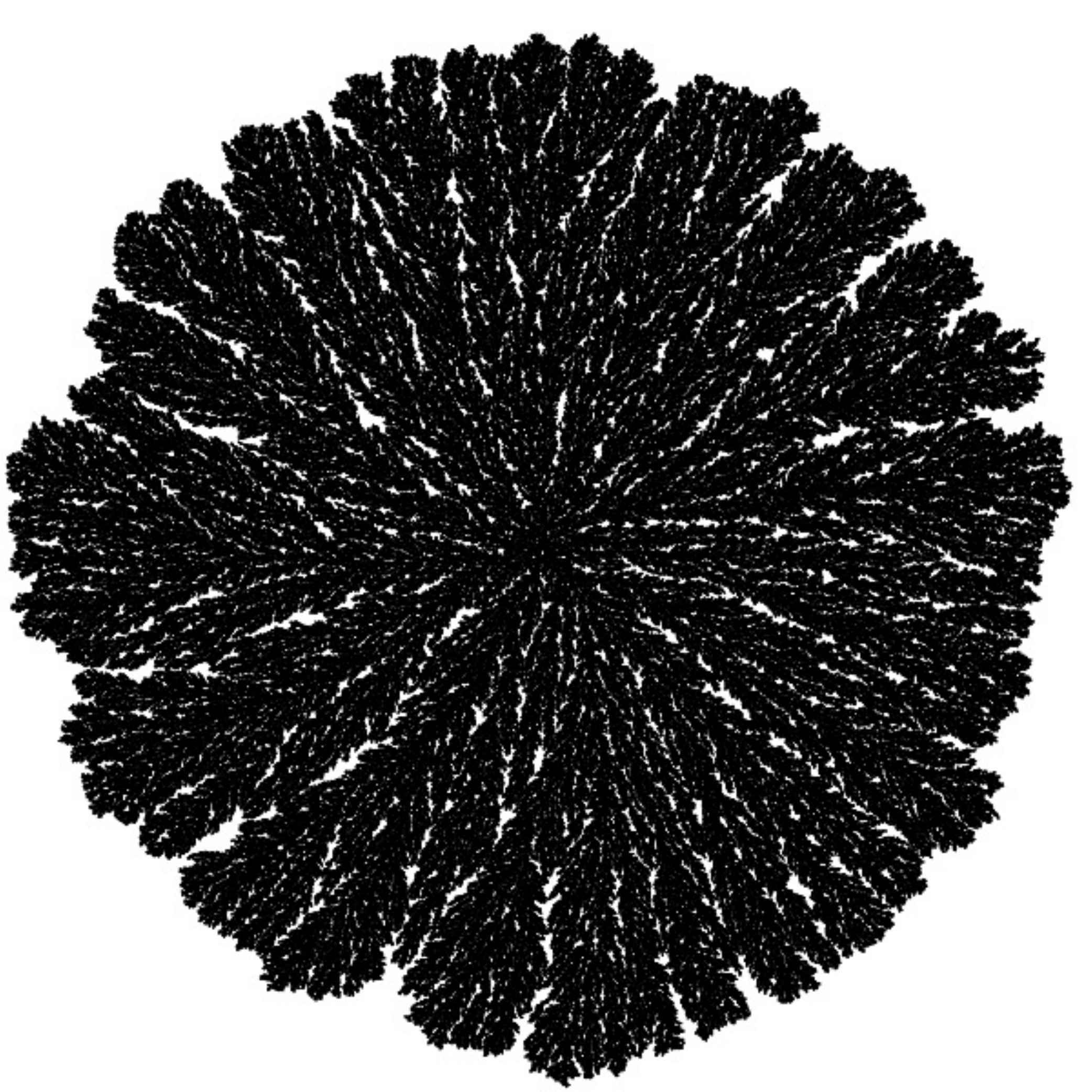}}}
\subfigure[~ $\delta_\theta = 20^\circ$]{\resizebox{3.2cm}{!}{\includegraphics{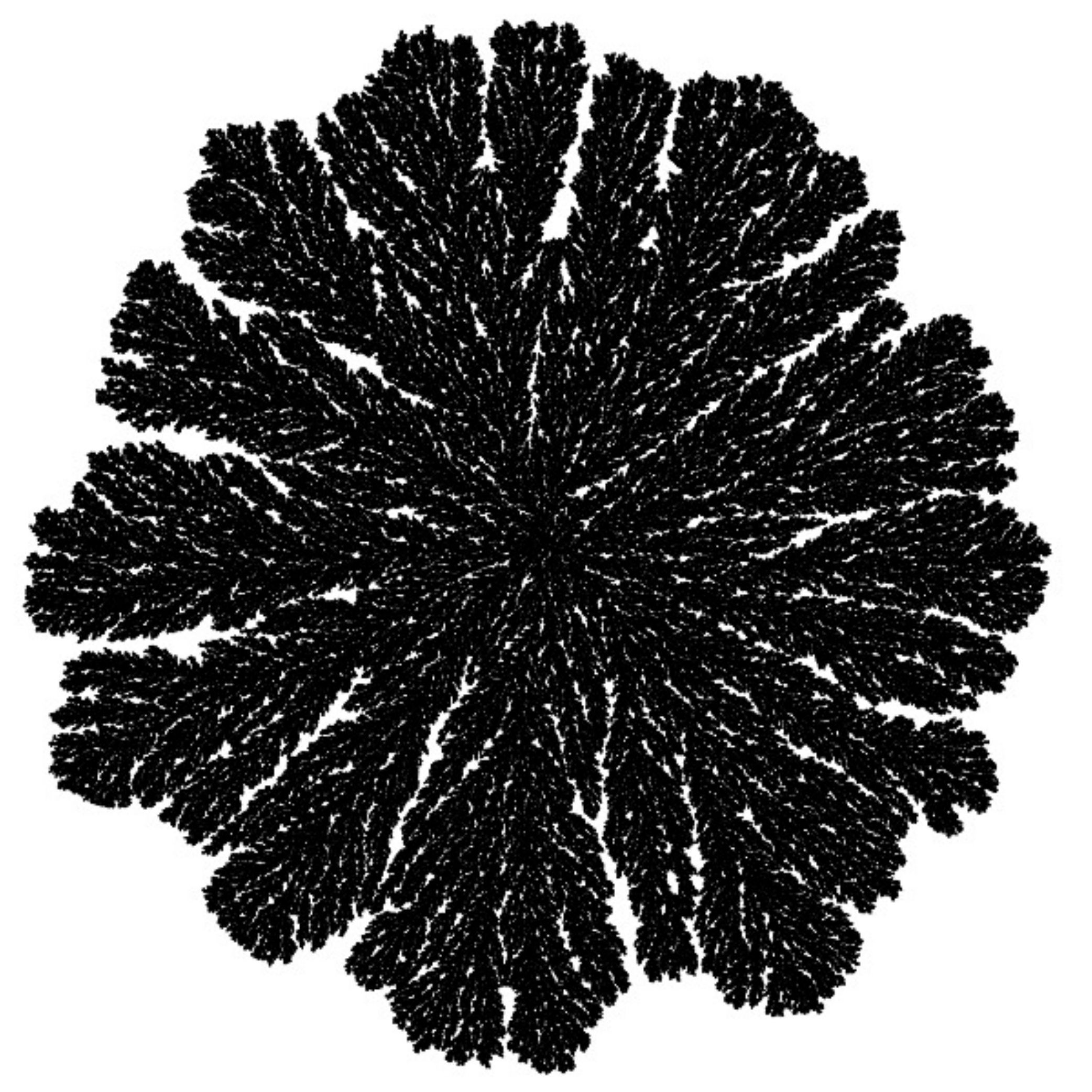}}}\\
\subfigure[~ $\delta_\theta = 40^\circ$]{\resizebox{3.2cm}{!}{\includegraphics{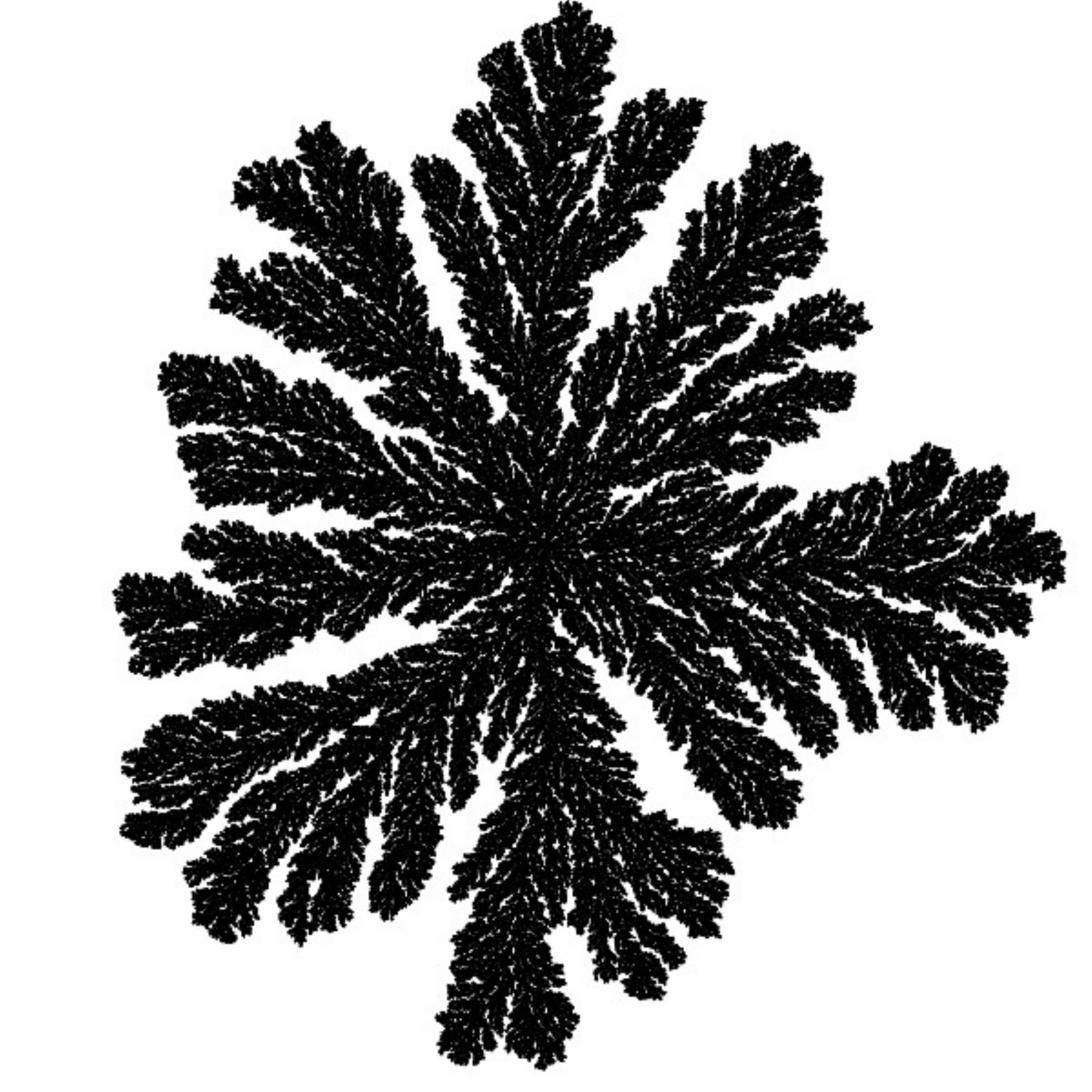}}}
\subfigure[~ $\delta_\theta = 120^\circ$]{\resizebox{3.2cm}{!}{\includegraphics{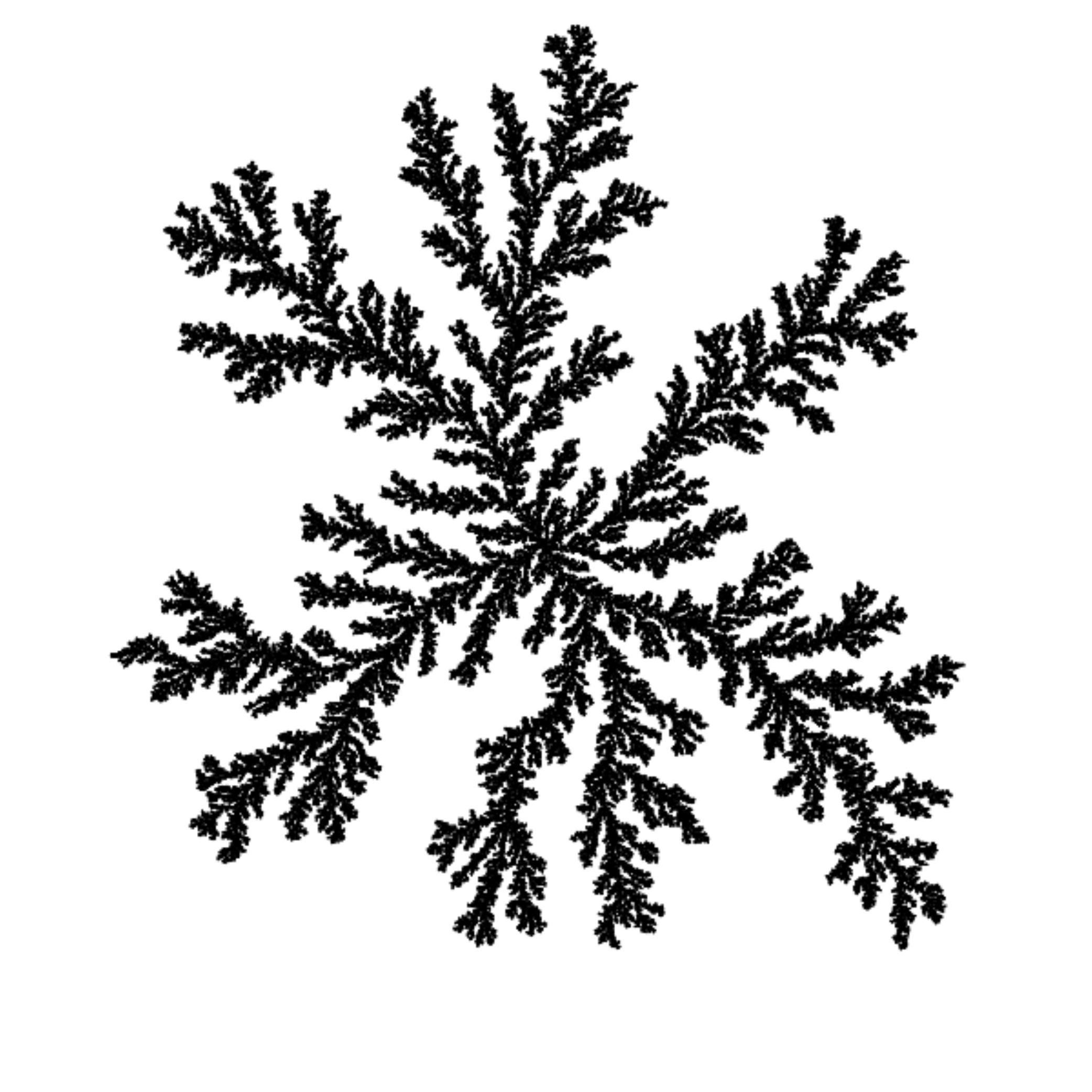}}}
\caption{\label{patterns} Clusters of a fixed size $r\approx 1000a$ obtained for distinct $\delta_\theta$ values.  The number of particles varies from $10^4$ for the most ramified to $10^6$ for the densest cluster. }                                                                                                                                                                                                                                                                                                                                     \end{center}
\end{figure}

This section is devoted to the simulations of a DLA model using the persistent
random walk defined by Eqs. (\ref{trajectory}) and (\ref{eq:phin}). Figure
\ref{stages} shows different stages of two clusters obtained for angular
openings $\delta_\theta = 40^\circ$ and $60^\circ$. In both cases, the change of
the cluster morphology is evident. At early stages, the aggregates have dense
branched morphologies (low density of inner voids) that resemble the BA patterns
while sparse branches, the hallmark of DLA clusters, are obtained after a
characteristic number of particles. This behaviour qualitatively confirms the
morphological crossover predicted with the trajectory properties:  the smaller
$\delta_\theta$ the longer the crossover time. Figure \ref{patterns} shows
clusters of the same size obtained with distinct $\delta_\theta$ values. The
simulations were stopped when the clusters reached a circle of radius $r=1000a$.
A transition from dense to ramified structures is observed and a fractal
dimension dependent on $\delta_\theta$ can  effectively be measured in agreement
with Ref. \cite{Huang}. However, it is important to notice that the dense
morphologies are transient and the patterns unavoidably become ramified with the
DLA fractal dimension ($D\sim 1.71$) for asymptotic large clusters.

The quantitative characterization of the crossover was done using the
mass-radius method \cite{Alves_PRE}. Clusters were grown until they cross a
circle of radius $5\times10^3a$. In order to perform statistical averages, 100
independent samples were simulated for each investigated $\delta_\theta$ value.
The mass-radius method consists in determining the number of particles inside a
circle of radius $r$ centred in the seed. In general, the mass scales as $M(r)
\sim r^D$, where $D$ is the fractal dimension. We have $D\simeq 1.71$ and $D\simeq 2$
for asymptotically large DLA and BA clusters, respectively
\cite{Alves_PRE,Ferreira_PRE}. 

\begin{figure}[t]
\begin{center}
{\resizebox*{7.0cm}{!}{\includegraphics*{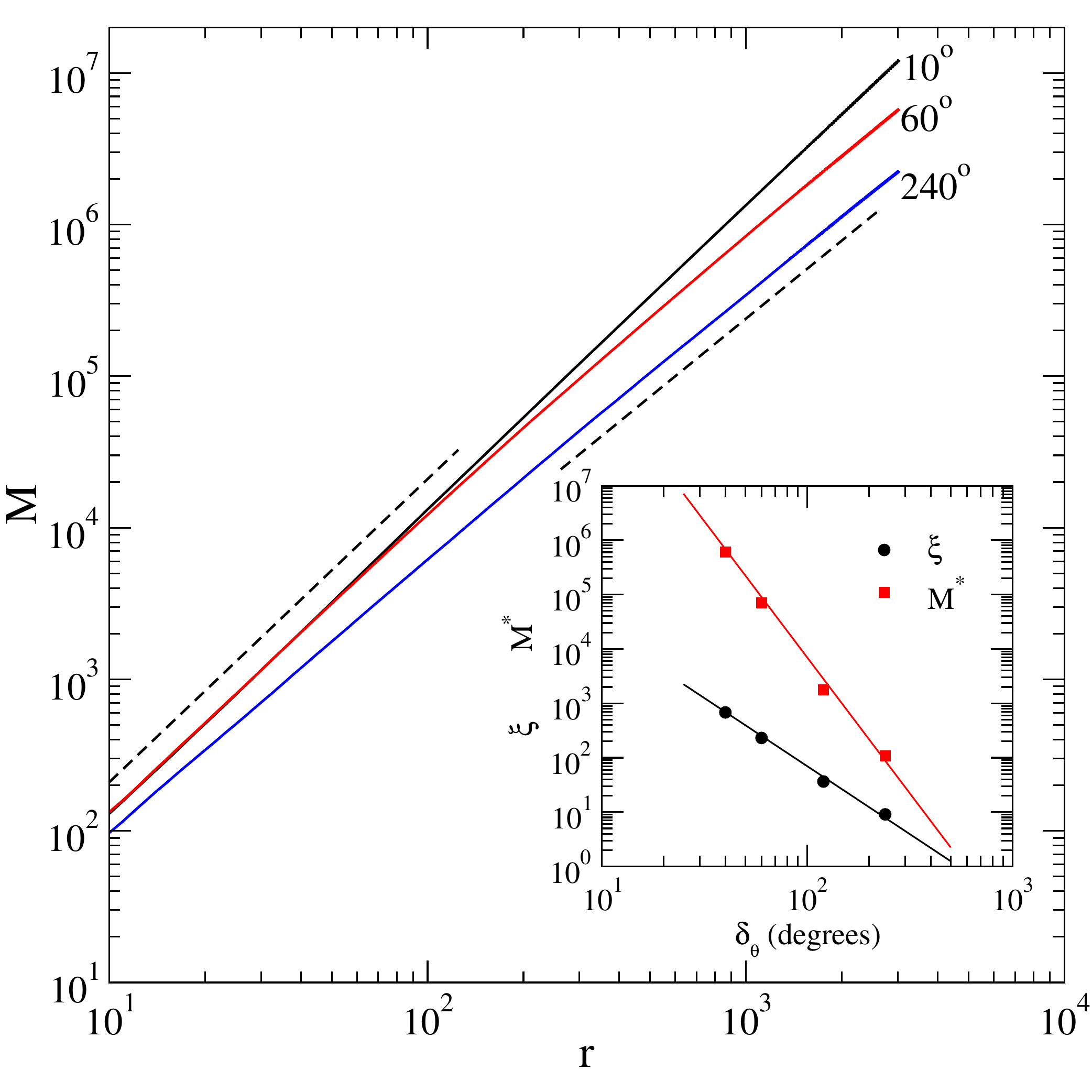}}}
\caption{\label{massradius} (Color on-line) Mass-radius curves for three
$\delta_\theta$ values. Dashed lines are power laws with exponents 2 (smaller
$r$) and 1.71 (larger $r$) included as  guides to the eyes. The inset shows the
mass $M^*$ and the characteristic radius $\xi$ at the crossover as functions of
$\delta_\theta$. Power law regressions with exponents $-5$ and $-2.5$ are also
shown in the inset. } 
\end{center}
\end{figure}

Figure \ref{massradius} shows the mass-radius curves corresponding to different
values of $\delta_\theta$. The curves have a crossover between the scaling
regimes $M\sim r^2$ at small radius and $M\sim r^{1.71}$
at large radius. The mass is described by a scaling form
\begin{equation}
\mbox{M}(r,\delta_\theta)= \delta_\theta^{-2z} 
                           f\left({\delta_\theta^{z}r}\right),
\label{mass_scaling}
\end{equation}
where $f(x) \sim x^{2}$ for $x\ll 1$ and  $f(x) \sim x^{1.71}$ for $x\gg 1$. This
scaling ansatz supposes that the crossover  radius scales as $\xi\sim
\delta_\theta^{-z}$. It also assumes a constraint  $M(\xi) \sim \xi^2$  due to
the clusters still are dense close to the crossover, explaining the exponent
$2z$ in Eq. (\ref{mass_scaling}). Determining the crossover radius as the intersection
between the scaling laws $M_1 \sim r^2$ and $M_2 \sim r^{1.71}$, we obtained
a scaling law very close to $\xi \sim \delta_\theta^{-2.5}$. We also evaluated the
mass at the crossover and the relation $M(\xi)\sim\xi^2$ was confirmed.  The
crossover mass and radius against $\delta_\theta$ are shown in the inset of Fig.
\ref{massradius}. 

\begin{figure}[t]
\begin{center}
\resizebox*{7cm}{!}{\includegraphics*{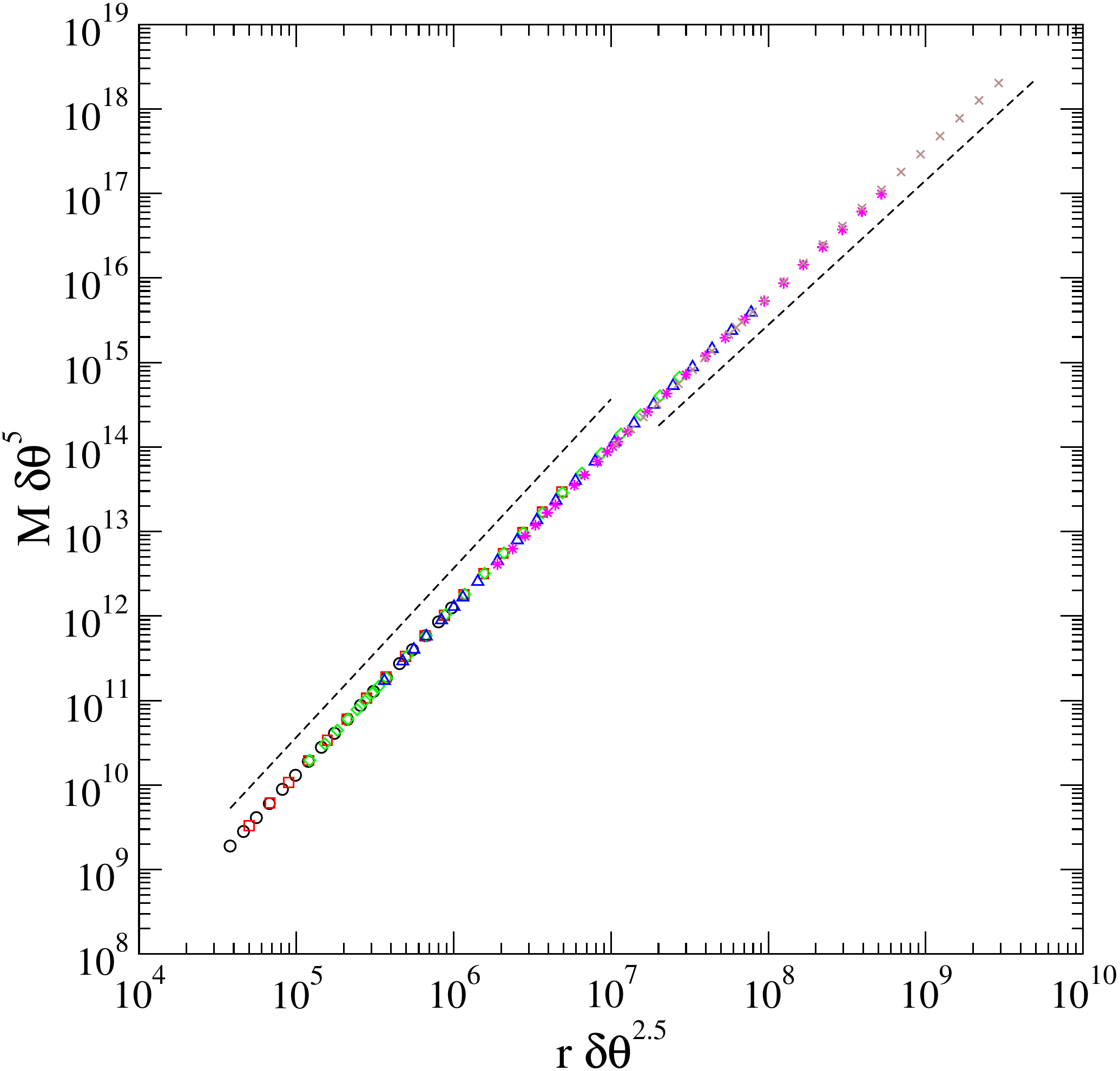}}
\end{center}
\caption{\label{colapso} (Colour on-line) Collapses of the mass-radius curves
for $\delta_\theta = 10^\circ$, $20^\circ$, $40^\circ$, $60^\circ$,  $120^\circ$, and $240^\circ$
using the scaling ansatz given by Eq. \eqref{mass_scaling}.
The solid lines are scaling laws $r^2$ and $r^{1.71}$ corresponding to the
asymptotic behaviour of the scaling function $f(x)$.}
\end{figure}

In Fig. \ref{colapso} the mass-radius curves are collapsed using the scaling
ansatz given by Eq. \eqref{mass_scaling} with the exponent $z=2.5$. As can be seen, an
excellent collapse onto the scaling function $f(x)$ is obtained. 
{For the general case with a drift quantified by a parameter $b$, the
mass against radius can be written as \cite{Alves_JCP}
\begin{equation}
\label{scaling_long}
M(r;b) = b^\alpha f\left( \frac{r}{b^\zeta} \right),
\end{equation}
where the scaling laws  $\xi\sim b^\zeta$ and $M(\xi)\sim b^\alpha$ are assumed
for the crossover radius and the corresponding mass, respectively. The scaling
function $f(x)$ yields the asymptotic regimes of $M(r)$: For a DLA to BA
transition, we have  $f(x)\sim x^{1.71}$ for $x\ll 1$ and $f(x)\sim x^{2}$ for
$x\gg 1$ while the scaling exponents are exchanged to  $f(x)\sim x^{2}$ for $x\ll 1$
and  $f(x)\sim x^{1.71}$ for $x\gg 1$ for a BA to DLA transition.} 
We can easily relate the scaling hypothesis \eqref{mass_scaling} with the scaling ansatz given
by Eq. \eqref{scaling_long} for the aggregation of particles performing long
steps of size $\ell=b$ . It was shown in Sec. \ref{sec:scaling} that the
characteristic length in a persistent random walk scales as
$\ell\sim\delta_\theta^{-2}$. Replacing $\delta_\theta\sim 1/\sqrt{\ell}$ in Eq.
\eqref{mass_scaling} and comparing with Eq. \eqref{scaling_long}, we find the
exponents $\alpha=z=2.5$ and $\zeta=z/2=1.25$. These exponents are consistent
with those obtained for long steps trajectories in Ref. \cite{Alves_JCP}
indicating a universality class. 

\section{Conclusions}
\label{conclusions}

In this work we investigated the scaling properties of a diffusion limited
aggregation (DLA) model where the particles follow locally persistent
trajectories. The bias is introduced by limiting the direction of a new step
into an angular opening $\delta_\theta$ in relation to the preceding step
direction. We show analytically and numerically that this trajectory is
effectively ballistic below a length scale $\ell\sim\delta_\theta^{-2}$ and
becomes random at large scales. This result improves the numerical estimate
of the scaling law reported formerly by Tojo and Argyrakis \cite{Tojo}.

Large scale simulations show that the aggregates undergo a morphological
transition from a ballistic aggregation (BA) regime at small sizes to a DLA-like
ramified morphology at asymptotically large scales. It is important to mention that
a previous analysis of this aggregation model \cite{Huang}, which was done in a
small cluster size limit, states a non-universal fractal dimension.
However, we have shown that the asymptotic behaviour of the clusters is described by
the DLA fractal dimension. Finally, the crossover between the BA and DLA scaling
regimes is related with the angular opening by $\xi\sim\delta_\ell^{-1.25}$
indicating a non-trivial relation between the characteristic size of the cluster 
in the crossover and the persistence length of the random walk. 

The persistent random walk has interesting features that can be associated to a
number of physical systems
\cite{Vilesov2007,Kolega,Wu2000,Vilesov2008,Peruani2007}. A small angular
opening may represent a diffusion in a medium where the scattering is weak
except by rare but no negligible fluctuations. An important example is the
wandering/aggregation of molecules in superfluid helium \cite{Vilesov2007} where
the particles may have a mean free path ranging from a molecule diameter to
micrometers depending on the proximity to the critical temperature. This kind of
diffusion is also observed in cell migration \cite{Dickinson,Kolega}, a
chemotactic process in which the cells diffuse following a gradient-mediated
chemical signalling. A relevant property of the trajectory is a characteristic
mean free path or persistence length that emerges naturally as we have
demonstrated with the central limit theorem \cite{vankampen}. In the cited
examples, the aggregation and the consequent formation of clusters were subjects
of recent experimental interest \cite{Vilesov2007,Vilela,Mendes2001}.
Consequently, the present theoretical analysis can be extended to future applied
investigations.

\section*{Acknowledgements}
This work was supported by the  Brazilian agencies FAPEMIG and CNPq.
SCF thanks the kind hospitality at the Departament de F\'{\i}sica i Enginyeria Nuclear/UPC.


\end{document}